\documentclass[letterpaper, 10 pt, conference]{ieeeconf}  

\IEEEoverridecommandlockouts                              
\usepackage{amsmath}
\usepackage{amssymb}
\usepackage{mathtools}
\usepackage[dvipsnames]{xcolor}
\usepackage{tikz}

\newtheorem{thm}{Theorem}
\newtheorem{lem}{Lemma}
\newtheorem{cor}{Corollary}
\newtheorem{defn}{Definition}
\newtheorem{rem}{Remark}
\newtheorem{assum}{Assumption}

\newcommand\copyrighttext{%
	\footnotesize \copyright 2022 IEEE. Personal use of this material is permitted. Permission from IEEE must be obtained for all other uses, in any current or future media, including reprinting/republishing this material for advertising or promotional purposes, creating new collective works, for resale or redistribution to servers or lists, or reuse of any copyrighted component of this work in other works.}
\newcommand\copyrightnotice{%
	\begin{tikzpicture}[remember picture,overlay]
		\node[anchor=south,yshift=10pt] at (current page.south) {\fbox{\parbox{\dimexpr\textwidth-\fboxsep-\fboxrule\relax}{\copyrighttext}}};
	\end{tikzpicture}%
}

\title{\LARGE \bf
On a Continuous-Time Version of Willems' Lemma
}

\author{Victor G. Lopez and Matthias A. Müller
\thanks{This work received funding from the European Research Council (ERC) under the European Union’s Horizon 2020 research and innovation programme (grant agreement No 948679).}
\thanks{V. G. Lopez and M. A. Müller are with the Leibniz University Hannover, Institute of Automatic Control, 30167 Hannover, Germany
        {\tt\small \{lopez,mueller\}@irt.uni-hannover.de}}%
}

\begin{document}

\maketitle
\thispagestyle{empty}
\pagestyle{empty}
\copyrightnotice

\begin{abstract}

In this paper, a method to represent every input-output trajectory of a continuous-time linear system in terms of previously collected data is presented. This corresponds to a continuous-time version of the well-known Willems' lemma. The result is obtained by sampling the continuous signals at regular intervals, and constructing Hankel-like structures that closely resemble their discrete-time counterparts. Then, it is shown how to use measured persistently excited data to design a time-varying vector of parameters that allows the generation of arbitrary piecewise differentiable trajectories. A class of input signals that satisfies the conditions for persistence of excitation is also provided. 

\end{abstract}

\section{INTRODUCTION}

In 2005, Willems and coauthors presented the seminal result \cite{WillemsRapMarDe2005}, where they show that every input-output trajectory of a controllable discrete-time linear time-invariant (DT-LTI) system can be expressed as a linear combination of time-shifts of a persistently excited trajectory measured from that system. This result, now known as Willems' lemma, allows describing the behavior of any controllable DT-LTI system in a data-based fashion. The implications of this method are of significant importance in system analysis and control, since it allows to directly employ measured data without an intermediate system identification procedure.

Willems' lemma, which was originally described in the behavioral setting in \cite{WillemsRapMarDe2005}, has been reproduced using the state-space framework in \cite{DePersisTes2020,vanWaardePerCamTes2020,BerberichAll2020}. In recent years, a large amount of extensions and applications of Willems' lemma in the context of data-driven system analysis and control has been published. For instance, it was used to design data-based stabilizing controllers in \cite{DePersisTes2020}. Robust controllers \cite{BerberichKocSchAll2020} and optimal controllers \cite{DePersisTes2020,DorflerTesDe2021} have also been obtained. Data-based predictive controllers were developed in \cite{CoulsonLygDoe2019,BerberichKoeMueAll2021}.  The data collection requirements in reinforcement learning algorithms have been improved \cite{LopezAlsMue2021}. In \cite{TuranFer2022,WolffLopMul2021}, the data-based state estimation problem was addressed. Besides these applications, the data-based representation of dynamical systems has been extended to linear time-varying systems \cite{VerhoekTotHaeKoc2021,NortmannMyl2021}, to stochastic systems \cite{PanOuFau2021}, and for classes of nonlinear systems \cite{BerberichAll2020,AlsaltiBerLopAllMue2021,RuedaSch2020}. A maximum likelihood approach was used to determine optimal system representations from noisy data in \cite{YinIanSmi2021}. Data-based verification of dissipativity was studied in \cite{RomerBerKohAll2019}. Moreover, Willems' lemma inspired the work on data informativity \cite{vanWaardeEisTreCam2020}, where the system properties that can be inferred from measured data are analyzed. In \cite{MarkovskyDor2021}, a comprehensive survey of the existing results using Willems' lemma can be found; all these results have been developed for discrete-time systems\footnote{Some papers \cite{DePersisTes2020,BerberichWilHerAll2021} address the data-based design of controllers for continuous-time systems. However, no data-based representation of continuous-time systems trajectories has been published.}.

Since most physical systems in practical applications have inherently continuous-time dynamics, it is of interest to obtain a data-based representation that describes their continuous-time behavior. Such a representation would allow the design of controllers directly for continuous-time systems. Furthermore, the solution of problems like, e.g., the data-based simulation of continuous-time systems could also be solved. Moreover, a data-based representation of continuous-time systems would also be of theoretical importance, since it would provide a potential tool for the analysis of the properties of a physical system without resorting to a discretization procedure.

In this paper, we present a continuous-time version of Willems' lemma, i.e., a method to generate the input-output trajectories of continuous-time linear time-invariant (CT-LTI) systems in a data-based fashion, analogous to the known results for DT-LTI systems. This enables data-based system analysis and control techniques, similar to the discrete-time case discussed above. The proposed method is based on the construction of time-varying Hankel matrices, obtained after partitioning the continuous-time signals into regular segments. In contrast to the discrete-time case, a time-varying parameter vector is designed such that its product with the data Hankel matrices spans system trajectories. This time-varying parameter vector is obtained by solving a suitable differential equation. The only condition for the spanned trajectories is that they must be generated by a piecewise differentiable input signal. We also provide conditions for the design of a suitable input, such that the data generated from the CT-LTI system is persistently excited.

This paper is organized as follows. In Section \ref{secprel}, we introduce relevant definitions and expressions that will be used throughout the paper. The proposed CT version of Willems' lemma is described in Section \ref{secmain}. Section \ref{secdisc} presents further discussions about the obtained results. Finally, our conclusions are presented in Section \ref{secconc}.

\section{PRELIMINARIES}
\label{secprel}

In this section, we present the notation and definitions that will be useful for our developments. The known Willems' lemma for DT systems is also briefly described.

\subsection{Definitions}

Let $z : [0,\bar N] \rightarrow \mathbb{R}^\sigma$, with $[0,\bar N] \subset \mathbb{R}$, denote a continuous-time signal of length $\bar N$. For convenience, let the signal length $\bar N$ be partitioned into $N$ regular intervals, such that $\bar N = NT$, with $N \in \mathbb{N}$ and for some $T \in \mathbb{R}_+$. Using the trajectory $z : [0,NT] \rightarrow \mathbb{R}^\sigma$, the time interval $T$, and two parameters $a,b \in \mathbb{N}_{0}$, $a < b \leq N-1$, we define the following matrix 
\begin{multline}
	H_{1,T}(z_{[a,b]}(t)) := \\
	\left[ \begin{array}{cccc}
		z(t+aT) & z(t+(a+1)T) & \cdots & z(t+bT) \end{array} \right]
	\label{hankrow}
\end{multline}
for $0 \leq t < T$. Thus, $H_{1,T}(z_{[a,b]}(t))$ is a time-varying matrix defined on the interval $t \in [0,T)$.

Of special interest is the following matrix, constructed by concatenating $L$ matrices of the form (\ref{hankrow}),
\begin{equation*}
	H_{L,T}(z_{[0,N-1]}(t)) := \left[ \begin{array}{c}
		H_{1,T}(z_{[0,N-L]}(t)) \\
		H_{1,T}(z_{[1,N-L+1]}(t)) \\
		\vdots \\
		H_{1,T}(z_{[L-1,N-1]}(t))
	\end{array} \right]
\end{equation*}
for $0 \leq t < T$.  Notice that, at each instant $t$, $H_{L,T}(z_{[0,N-1]}(t))$ corresponds to a Hankel matrix for the discrete-time sequence obtained by sampling the signal $z : [0,NT] \rightarrow \mathbb{R}^\sigma$ at every $T$ time units, starting at time $t$.

\subsection{State trajectories of continuous-time systems}

Consider a CT-LTI system
\begin{align}
	\begin{split}
		\dot x(t) & = Ax(t) + Bu(t), \\
		y(t) & = Cx(t) + Du(t),
	\end{split}
\label{linsys}
\end{align}
where $x \in \mathbb{R}^n$, $u \in \mathbb{R}^m$ and $y \in \mathbb{R}^p$ are the state, input and output vectors of the system, respectively. Given the initial conditions $x(0)$ and an input signal $u : [0,t] \rightarrow \mathbb{R}^m$, the state of system (\ref{linsys}) at time $t$ is given by
\begin{equation}
	x(t) = e^{At} x(0) + \int_0^t e^{A(t - \tau)} B u(\tau) d\tau.
	\label{xtr}
\end{equation}

In the next section, we will take advantage of different expressions of the state trajectory obtained from (\ref{xtr}). In particular, for any nonnegative scalars $i$ and $T$, we can write
\begin{align}
	x(t + iT) & = e^{At} x(iT) + \int_{iT}^{t+iT} e^{A(t + iT - \tau)} B u(\tau) d\tau \nonumber \\[0.2mm]
	& = e^{At} x(iT) + \int_{0}^{t} e^{A(t - \bar \tau)} B u(\bar \tau + iT) d \bar \tau,
	\label{xittr}
\end{align}
where the second equation is obtained after the change of variables $\bar \tau = \tau - iT$.

Moreover, evaluating (\ref{xittr}) for $t = T$ results in
\begin{align}
	x((i+1)T) & = e^{AT} x(iT) + \! \int_{0}^{T} \! \! \! e^{A(T - \tau)} B u(\tau + iT) d \tau.
	\label{xit1tr}
\end{align}

\subsection{Discrete-time Willems' lemma}

The developments in this paper to obtain data-based representations of continuous-time system trajectories use the known results for discrete-time systems as a starting point. In order to better emphasize the similarities and differences between both cases, we briefly state here the well-known discrete-time Willems' lemma.

The Hankel matrix of depth $L$ of a discrete-time sequence $\{ z_k \}_{k=0}^{M-1} = \{z_0, \, z_1,\, \ldots,\, z_{M-1} \}$, $z_k \in \mathbb{R}^\sigma$, is given by
\begin{equation*}
	H_L(z) := \left[ \begin{array}{cccc}
		z_0 & z_1 & \cdots & z_{M-L} \\
		z_1 & z_2 & \cdots & z_{M-L+1} \\
		\vdots & \vdots & \ddots & \vdots \\
		z_{L-1} & z_{L} & \cdots & z_{M-1}
	\end{array} \right].
\end{equation*}
The discrete sequence $\{ z_k \}_{k=0}^{M-1}$ is said to be \textit{persistently exciting} of order $L$ if its Hankel matrix of depth $L$ has full row rank, i.e., 
\begin{equation}
	\text{rank} ( H_L(z) ) = \sigma L.
	\label{dtpe}
\end{equation}

Willems' lemma shows that, when a sufficiently exciting input is applied to a controllable DT system of the form
\begin{align}
	\begin{split}
		x_{k+1} & = Ax_k + Bu_k, \\
		y_k & = Cx_k + Du_k,
	\end{split}
\label{dtsyst}
\end{align}
the data collected from it contains enough information to describe all possible input-output trajectories that the system can generate. This result is stated as follows.

\begin{thm}[{\cite{WillemsRapMarDe2005}}]
	\label{thmdtwill}
	Consider the system (\ref{dtsyst}) and let the pair $(A,B)$ be controllable. Let the sequence $\{ u_k \}_{k=0}^{M-1}$ be persistently exciting of order $L+n$, and let $\{ x_k \}_{k=0}^{M-1}$ and $\{ y_k \}_{k=0}^{M-1}$ be the corresponding state and output trajectories of (\ref{dtsyst}), respectively. Then, the full row rank condition 
	\begin{equation}
		\text{rank} \left( \left[ \begin{array}{c}
			H_L(u) \\ H_1(\{ x_k \}_{k=0}^{M-L})
		\end{array} \right] \right) = Lm + n
		\label{willemrank}
	\end{equation}
	is satisfied. Moreover, any pair of sequences $\{ \bar u_k \}_{k=0}^{L-1}$, $\{ \bar y_k \}_{k=0}^{L-1}$ is an input-output trajectory of (\ref{dtsyst}) if and only if there exists a vector $\alpha$ such that
	\begin{equation}
		\left[ \begin{array}{c}
			H_L(u) \\ H_L(y)
		\end{array} \right] \alpha = \left[ \begin{array}{c}
		\bar U \\ \bar Y
	\end{array} \right] 
		\label{dtwillem}
	\end{equation}
where $\bar U = [\bar u_0^\top \,\, \cdots \,\, \bar u_{L-1}^\top]^\top$ and $\bar Y = [\bar y_0^\top \,\, \cdots \,\, \bar y_{L-1}^\top]^\top$.
\end{thm}

In the following section, we propose a method for CT systems which preserves the general structure of (\ref{dtwillem}), but fundamental differences arise in order to address the particular properties of continuous-time signals.

\section{WILLEMS' LEMMA FOR CONTINUOUS-TIME SYSTEMS}
\label{secmain}

In this section, the main results of the paper are presented. As in the DT Willems' lemma, our results require the a priori collection of persistently excited data from the system (\ref{linsys}). Thus, we begin our analysis by studying the conditions for persistence of excitation in continuous-time systems.

\subsection{Persistence of excitation for CT systems}

As described above, in \cite{WillemsRapMarDe2005} a persistently exciting (PE) input for DT systems is defined from the full rank condition (\ref{dtpe}). The key point of this definition is that, under controllability conditions, the input-state trajectory obtained from the application of this input fulfills the full rank condition (\ref{willemrank}). This property allows the subsequent generation of any trajectory of the DT system (\ref{dtsyst}) by means of (\ref{dtwillem}).

In this subsection, we define suitable persistence of excitation conditions for continuous-time systems. In particular, we define \emph{persistently excited data} such that a similar rank property as in (\ref{willemrank}) holds. Subsequently, we present a class of input signals that guarantee this condition.

Thus, we use in this paper the following definition of a PE data set (compare with Definition 1 in \cite{BerberichKoeMulAll2022}).

\begin{defn}
	\label{defpe}
	Consider the system (\ref{linsys}). The data $u : [0,NT] \rightarrow \mathbb{R}^m$, $x : [0,NT] \rightarrow \mathbb{R}^n$, $T > 0$, $N \in~\mathbb{N}$, is said to be persistently excited of order $L+n$ if the full row rank condition
	\begin{equation}
		\text{rank} \left( \left[ \begin{array}{c}
			 H_{L,T}(u_{[0,N-1]}(t)) \\ H_{1,T}(x_{[0,N-L]}(t))
		\end{array} \right] \right) = Lm + n
	\label{pecond}
	\end{equation}
is satisfied for all $0 \leq t <T$, where $x(t)$ is the system state resulting from the application of $u(t)$.
\end{defn}

\begin{rem}
	Notice that, for the matrix on the left-hand side of (\ref{pecond}) to be full row rank, a necessary condition is that the number of columns is at least as large as the number of rows, implying that $N \geq L(m+1) +n -1$. Since the length of the collected trajectories is $NT$, then a large $N$ implies that either the trajectory length is large enough (for fixed $T$), or $T$ is small enough (for fixed trajectory length). 
\end{rem}

Definition \ref{defpe} implies that the PE condition of a data set depends on the system (\ref{linsys}) under study, its initial conditions, and the input applied for data generation. However, it is possible to design input signals that guarantee the satisfaction of (\ref{pecond}) independently of the system matrices $(A,B,C,D)$ and the initial conditions, as long as the pair $(A,B)$ is controllable. This is proven in the following lemma, where we take advantage of the known properties of discrete-time PE inputs to design a suitable piecewise constant input for CT systems. This lemma requires the following mild assumption on the selection of the time interval $T$.

\begin{assum}
	\label{assumt}
	The time interval $T$ is such that
	\begin{equation*}
		T \neq \frac{2 \pi k}{| \mathcal{I}_m(\lambda_i - \lambda_j) |}, \qquad \forall k \in \mathbb{Z},
	\end{equation*}
where $\lambda_i$ and $\lambda_j$ are any two eigenvalues of matrix $A$ in (\ref{linsys}), and $\mathcal{I}_m(\cdot)$ is the imaginary part of a complex number.
\end{assum}

\begin{rem}
	Even with no knowledge of the system model (\ref{linsys}), Asumption \ref{assumt} is hardly restrictive. In fact, the values of $T$ that make this assumption fail form a set of measure zero and are, hence, unlikely to be encountered in practice.
\end{rem}

The first main result of the paper is stated as follows.

\begin{lem}
	\label{lempe}
	Consider system (\ref{linsys}), let the pair $(A,B)$ be controllable, and let Assumption \ref{assumt} hold. Consider a sequence of constant vectors $\{ \mu_i \}_{i=0}^{N-1}$, $\mu_i \in \mathbb{R}^m$, such that the matrix
	\begin{equation*}
		H_{L+n}(\mu) = \left[ \begin{array}{cccc}
			\mu_0 & \mu_1 & \cdots & \mu_{N-L-n} \\
			\mu_1 & \mu_2 & \cdots & \mu_{N-L-n+1} \\
			\vdots & \vdots & \ddots & \vdots \\
			\mu_{L+n-1} & \mu_{L+n} & \cdots & \mu_{N-1}
		\end{array} \right]
	\end{equation*} 
	has full row rank, i.e.,
	\begin{equation}
		\text{rank} \left( H_{L+n}(\mu) \right) = (L+n)m.
		\label{rankdt}
	\end{equation} 
	Then, the data $u : [0,NT] \rightarrow \mathbb{R}^m$, $x : [0,NT] \rightarrow \mathbb{R}^n$ is persistently excited of order $L+n$ in the sense of Definition~\ref{defpe} if the input $u$ is selected as the piecewise constant signal $u(t + iT) = \mu_i$ for all $0 \leq t < T$, $i=0,\ldots,N-1$.
	
\end{lem}
\begin{proof}
Since $u$ is piecewise constant, the state expression in (\ref{xit1tr}) can be written as
\begin{align*}
	x((i+1)T) & = e^{AT} x(iT) + \int_{0}^{T} e^{A(T - \tau)} B u(iT) d \tau \\
	& = e^{AT} x(iT) + \int_{0}^{T} e^{A(T - \tau)} d \tau B \mu_i.
\end{align*}
This is an exact discretization expression, where the matrices $\left( e^{AT},\, \int_{0}^{T} e^{A(T - \tau)} d \tau B \right)$ are known to form a controllable pair by controllability of system (\ref{linsys}) and the condition in Assumption \ref{assumt} \cite{Chen1999}. 

Now, notice that the condition (\ref{rankdt}) implies that the sequence $\{ \mu_0,\, \ldots,\, \mu_{N-1} \}$ fulfills the discrete-time conditions for persistence of excitation of order $L+n$ (see (\ref{dtpe})). Since the exactly discretized system $\left( e^{AT},\, \int_{0}^{T} e^{A(T - \tau)} d \tau B \right)$ is controllable and (\ref{rankdt}) holds, Theorem \ref{thmdtwill} guarantees that the matrix
\begin{multline}
	\left[ \begin{array}{c}
		H_{L}(\mu) \\ H_{1,T}(x_{[0,N-L]}(0))
	\end{array} \right] = \\
	\left[ \begin{array}{cccc}
		\mu_0 & \mu_1 & \cdots & \mu_{N-L} \\
		\mu_1 & \mu_2 & \cdots & \mu_{N-L+1} \\
		\vdots & \vdots & \ddots & \vdots \\
		\mu_{L-1} & \mu_L & \cdots & \mu_{N-1} \\
		x(0) & x(T) & \cdots & x((N-L)T)
	\end{array} \right]
	\label{tzero}
\end{multline}
has full row rank. By definition of the signal $u$ in the statement of the lemma, we have that $H_{L,T}(u_{[0,N-1]}(t)) = H_L(\mu)$ for all $0 \leq t < T$. This directly implies that condition (\ref{pecond}) is satisfied for $t=0$.

Finally, we show by contradiction that (\ref{pecond}) holds also for $0 < t < T$. Assume that there exist vectors $\xi \in \mathbb{R}^n$ and $\eta_j \in \mathbb{R}^m$, $j=1,\ldots,L$, not all zeros, such that
\begin{multline}
	\left[ \begin{array}{cccc}
		\eta_1^\top & \cdots & \eta_L^\top & \xi^\top
	\end{array} \right] \times \\
	\left[ \begin{array}{ccc}
		\mu_0 & \cdots & \mu_{N-L} \\
		\vdots & \ddots & \vdots \\
		\mu_{L-1} & \cdots & \mu_{N-1} \\
		x(t) & \cdots & x(t + (N-L)T)
	\end{array} \right] = 0
	\label{cont}
\end{multline}
Now, notice that we can use the expression (\ref{xittr}) and the fact that $x(t+iT)$ is the result of applying the constant input $\mu_i$ for $t$ time units from the initial condition $x(iT)$, to write
\begin{multline*}
	\left[ \begin{array}{ccccc}
		\int_0^t e^{A(t - \tau)} d \tau B & 0 & \cdots & 0 & e^{At}
	\end{array} \right] \times \\
	\left[ \begin{array}{ccc}
		\mu_0 & \cdots & \mu_{N-L} \\
		\mu_1 & \cdots & \mu_{N-L+1} \\
		\vdots & \ddots & \vdots \\
		\mu_{L-1} & \cdots & \mu_{N-1} \\
		x(0) & \cdots & x((N-L)T)
	\end{array} \right] \\
	= \left[ \begin{array}{ccc}
		x(t) & \cdots & x(t + (N-L)T)
	\end{array} \right].
\end{multline*}

Therefore, 
\begin{multline*}
	\left[ \begin{array}{ccccc}
		\xi^\top \int_0^t e^{A(t - \tau)} d \tau B + \eta_1^\top & \eta_2^\top & \cdots & \eta_L^\top & \xi^\top e^{At}
	\end{array} \right] \times \\
	\left[ \begin{array}{ccc}
		\mu_0 & \cdots & \mu_{N-L} \\
		\mu_1 & \cdots & \mu_{N-L+1} \\
		\vdots & \ddots & \vdots \\
		\mu_{L-1} & \cdots & \mu_{N-1} \\
		x(0) & \cdots & x((N-L)T)
	\end{array} \right] \\
	= \left[ \begin{array}{cccc}
		\eta_1^\top & \cdots & \eta_L^\top & \xi^\top
	\end{array} \right] \times \\
	\left[ \begin{array}{ccc}
		\mu_0 & \cdots & \mu_{N-L} \\
		\vdots & \ddots & \vdots \\
		\mu_{L-1} & \cdots & \mu_{N-1} \\
		x(t) & \cdots & x(t + (N-L)T)
	\end{array} \right] \stackrel{(\ref{cont})}{=} 0
\end{multline*}
Since $\left[ \begin{array}{c}  H_{L}(\mu) \\ H_{1,T}(x_{[0,N-L]}(0)) \end{array} \right]$ in (\ref{tzero}) was shown to have full row rank, it follows that
\begin{equation*}
	\left[ \begin{array}{ccccc}
		\xi^\top \int_0^t e^{A(t - \tau)} d \tau B + \eta_1^\top & \eta_2^\top & \cdots & \eta_L^\top & \xi^\top e^{At}
	\end{array} \right] = 0.
\end{equation*}
This directly implies $\eta_2 = \eta_3 = \cdots = \eta_L =0$. Moreover, $e^{At}$ is nonsingular at every $t$, and therefore $\xi^\top e^{At} = 0$ implies $\xi = 0$. Finally, substituting this result in $\xi^\top \int_0^t e^{A(t - \tau)} d \tau B + \eta_1^\top = 0$ yields $\eta_1 = 0$. This contradicts the assumption that the vector $\left[ \begin{array}{cccc}
	\eta_1^\top & \cdots & \eta_L^\top & \xi^\top
\end{array} \right]$ in (\ref{cont}) was nonzero, completing the proof.
\end{proof}

Lemma \ref{lempe} shows that an input that is PE for discrete-time systems also allows the generation of PE data from continuous-time systems, provided that the time interval $T$ satisfies Assumption~\ref{assumt}. Assumption \ref{assumt} guarantees that the exactly discretized system that corresponds to applying a piecewise constant PE input to system (\ref{linsys}) is controllable \cite{Chen1999}. Notice also that Assumption~\ref{assumt} is a sufficient but not necessary condition for Lemma~\ref{lempe} to hold. Designing more general (non-piecewise constant) inputs that satisfy the PE condition for CT systems is a subject for future research.

In the following subsection, measured persistently excited data is used to generate system trajectories from CT systems.

\subsection{Willems' lemma for continuous-time systems}

In this subsection, we obtain a data-based representation of the trajectories of a controllable system (\ref{linsys}), using a measured persistently excited trajectory. Here, we consider the case where the generated input signal is differentiable. In the next section, we extend this result to piecewise differentiable signals.

Different from the discrete-time case, arbitrary continuous signals cannot be obtained using a constant vector $\alpha$ as in (\ref{dtwillem}). Thus, we allow the vector $\alpha(t)$ to be a function of time. In particular, let $\alpha(t)$ satisfy the differential equation with initial condition constraints presented in the following lemma.

\begin{lem}
	\label{lemdifeq}
	Consider the system (\ref{linsys}). Let $u : [0,NT] \rightarrow \mathbb{R}^m$, $x : [0,NT] \rightarrow \mathbb{R}^n$, $T > 0$, $N \in \mathbb{N}$, be persistently excited data of order $n+1$, where $u$ is continuously differentiable for all $t \in [0,NT]$ with $t \neq iT$, $i=1,\ldots,N-1$, and continuously differentiable from the right at the instants $t = iT$. Moreover, consider an arbitrary continuously differentiable signal $\bar u : [0,T] \rightarrow \mathbb{R}^m$ and let $\bar x(0) \in \mathbb{R}^n$ be an arbitrary vector. Then, there exists a solution $\alpha(t)$ of the differential equation with initial condition constraint
	\begin{multline}
		\left[ \begin{array}{c}
			H_{1,T}(u_{[0,N-1]}(t)) \\ H_{1,T}(x_{[0,N-1]}(t))
		\end{array} \right] \dot \alpha(t) = \\
		- \left[ \begin{array}{c}
			H_{1,T}(\dot u_{[0,N-1]}(t)) \\ 0
		\end{array} \right] \alpha(t) + \left[ \begin{array}{c}
			\dot{\bar{u}}(t) \\ 0
		\end{array} \right],
		\label{difeqa}
	\end{multline}
	\begin{equation}
		\left[ \begin{array}{c}
			H_{1,T}(u_{[0,N-1]}(0)) \\ H_{1,T}(x_{[0,N-1]}(0))
		\end{array} \right] \alpha(0) = \left[ \begin{array}{c}
			\bar{u}(0) \\ \bar x(0)
		\end{array} \right],
		\label{inicoa}
	\end{equation}
	for $0 \leq t < T$.
\end{lem}
\begin{proof}
	By persistency of excitation of the collected input-state data, $\left[ \begin{array}{c} H_{1,T}(u_{[0,N-1]}(t)) \\ H_{1,T}(x_{[0,N-1]}(t)) \end{array} \right]$ has full row rank for all $0 \leq t < T$ (compare (\ref{pecond})). Let $\Theta(t)$ be the Moore-Penrose pseudoinverse of this matrix. Hence, if the ordinary differential equation (ODE) 
		\begin{equation}
			\dot \alpha(t) = - \Theta(t) \left[ \begin{array}{c}
				H_{1,T}(\dot u_{[0,N-1]}(t)) \\ 0
			\end{array} \right] \alpha(t) + \Theta(t) \left[ \begin{array}{c}
				\dot{\bar{u}}(t) \\ 0
			\end{array} \right],
			\label{odea}
		\end{equation}
	with initial conditions
	\begin{equation*}
		\alpha(0) = \Theta(0) \left[ \begin{array}{c}
			\bar{u}(0) \\ \bar x(0)
		\end{array} \right],
	\end{equation*}
	has a solution $\alpha(t)$, then this solution must also satisfy (\ref{difeqa})-(\ref{inicoa}). Now, the solution of a differential equation of the form (\ref{odea}) is known to exist and be unique for arbitrary initial conditions if the elements of the matrix $\Theta(t) \left[ \begin{array}{c} H_{1,T}(\dot u_{[0,N-1]}(t)) \\ 0 \end{array} \right]$ are continuous in $t$ \cite{Chen1999}. This is the case since (i) $u(t)$ is continuously differentiable for all $t \in [0,NT]$ with $t \neq iT$, $i=1,\ldots,N-1$, and (ii) continuity of the Moore-Penrose pseudoinverse for full-rank matrices has been shown in \cite{Stewart1969}.
\end{proof}

\begin{rem}
	Although we expressed (\ref{difeqa}) as an implicit differential equation, in practice a solution $\alpha(t)$ can be obtained by solving the ODE (\ref{odea}) (see Section \ref{subsecprac}). Moreover, notice that (\ref{difeqa}) is presented in a general form, where $\dot u(t)$ can be nonzero. However, if a piecewise constant input as described in Lemma \ref{lempe} is used, then (\ref{difeqa}) becomes
	\begin{equation*}
		\left[ \begin{array}{c}
			H_{1,T}(u_{[0,N-1]}(t)) \\ H_{1,T}(x_{[0,N-1]}(t))
		\end{array} \right] \dot \alpha(t) = \left[ \begin{array}{c}
			\dot{\bar{u}}(t) \\ 0
		\end{array} \right].
	\end{equation*}
\end{rem}

The following theorem corresponds to the continuous-time version of Willems' lemma proposed in this paper

\begin{thm}
	\label{thmctwill}
	 Consider system (\ref{linsys}). Let $u : [0,NT] \rightarrow \mathbb{R}^m$, $x : [0,NT] \rightarrow \mathbb{R}^n$, $T > 0$, $N \in \mathbb{N}$, be persistently excited data of order $n+1$, where $u$ is continuously differentiable for all $t \in [0,NT]$ with $t \neq iT$, $i=1,\ldots,N-1$, and continuously differentiable from the right at the instants $t = iT$, and let $y : [0,NT] \rightarrow \mathbb{R}^p$ be the corresponding output trajectory of the system. Then, any signals $\bar u : [0,T] \rightarrow \mathbb{R}^m$, $\bar y : [0,T] \rightarrow \mathbb{R}^p$, where $\bar u$ is continuously differentiable, are an input-output trajectory of (\ref{linsys}) corresponding to some initial condition $\bar x(0)$ if and only if there exists a continuously differentiable vector $\alpha(t)$ such that the equations (\ref{difeqa}), (\ref{inicoa}) and
	 \begin{equation}
	 	\left[ \begin{array}{c}
	 		H_{1,T}(u_{[0,N-1]}(t)) \\ H_{1,T}(y_{[0,N-1]}(t))
	 	\end{array} \right] \alpha(t) = \left[ \begin{array}{c}
	 		\bar u(t) \\ \bar y(t)
	 	\end{array} \right]
	 	\label{ctwillem}
	 \end{equation}
	 hold for $0 \leq t < T$.
\end{thm}
\begin{proof}
	Without risk of confusion, in this proof we simplify our notation as $H(z(t)) := H_{1,T}(z_{[0,N-1]}(t))$ for any signal $z$ of length $NT$.
	
	\textit{Sufficiency}. First notice that, given the state signal $x$, the time derivative of the product $H(x(t)) \alpha(t)$ is given by
	\begin{align}
		\frac{d}{dt} \left( H(x(t)) \alpha(t) \right) = & H(\dot x(t)) \alpha(t) + H(x(t)) \dot \alpha(t) \nonumber \\
		= & \left( A H(x(t)) + B H(u(t)) \right) \alpha(t) \nonumber \\
		& + H(x(t)) \dot \alpha(t)
		\label{hdyn}
	\end{align}
	Now, assume that $\alpha(t)$ satisfies (\ref{difeqa})-(\ref{inicoa}) and (\ref{ctwillem}), and define $\bar x(t) := H(x(t)) \alpha(t)$. Then, from (\ref{ctwillem}) and (\ref{hdyn}) we have
	\begin{align*}
		\dot{\bar{x}}(t) & = A H(x(t)) \alpha(t) + B H(u(t)) \alpha(t) + H(x(t)) \dot \alpha(t) \\
		& = A \bar x(t) + B \bar u(t) + H(x(t)) \dot \alpha(t)
	\end{align*}
	From (\ref{difeqa}), we have that $H(x(t)) \dot \alpha(t) = 0$, which implies $\dot{\bar{x}} = A \bar x(t) + B \bar u(t)$ and, therefore, $\bar x(t)$ corresponds to the state trajectory of system (\ref{linsys}) with input $\bar u(t)$ and initial condition $\bar x(0) = H(x(0)) \alpha(0)$. Finally, from (\ref{ctwillem}) and the definition of the output in (\ref{linsys}), we can write
	\begin{align*}
		\bar y(t) & = H(y(t)) \alpha(t) = CH(x(t)) \alpha(t) + D H(u(t)) \alpha(t) \\
		& = C \bar x(t) + D \bar u(t).
	\end{align*}
	Therefore, $\bar y(t)$ corresponds to the output of system (\ref{linsys}).

\textit{Necessity}. Assume that $\bar u$, $\bar y$ is an input-output trajectory of the system. Since the conditions of Lemma \ref{lemdifeq} are satisfied, there exists a solution $\alpha(t)$ to (\ref{difeqa})-(\ref{inicoa}). Now, we show that this $\alpha(t)$ satisfies (\ref{ctwillem}). From (\ref{difeqa}), we observe that $H(u(t)) \dot \alpha(t) = -H(\dot u(t)) \alpha(t) + \dot{\bar{u}}(t)$. Together with (\ref{inicoa}), this implies that $\alpha(t)$ satisfies $H(u(t)) \alpha(t) = \bar u(t)$, and from (\ref{difeqa}) we have that $H(x(t)) \dot \alpha(t) = 0$. Using these expressions, the state trajectory that corresponds to the input $\bar u(t)$ satisfies
	\begin{align*}
		\dot{\bar{x}}(t) & = A \bar x(t) + B \bar u(t) \\
		& = A \bar x(t) + B H(u(t)) \alpha (t) + H(x(t)) \dot \alpha(t)
	\end{align*}
Comparing this expression to (\ref{hdyn}), and since $H(x(0)) \alpha(0) = \bar x(0)$ holds from (\ref{inicoa}), we conclude that $\bar x(t) = H(x(t)) \alpha(t)$. Finally, the output trajectory $\bar y(t)$ is given by
\begin{align*}
	\bar y(t) & = C \bar x(t) + D \bar u(t) = C H(x(t)) \alpha(t) + D H(u(t)) \alpha(t) \\
	& = H(y(t)) \alpha(t). 
\end{align*}
as claimed.
\end{proof}

Theorem \ref{thmctwill} provides sufficient and necessary conditions to represent input-output trajectories of length $T$ of a CT-LTI system using a previously collected persistently excited trajectory. In the following section, we discuss relevant features and possible extensions of this result.

\section{DISCUSSION}
\label{secdisc}

\subsection{On trajectories with piecewise continuous inputs}

The only restriction on the system trajectories that can be represented in a data-based fashion via Theorem \ref{thmctwill} is that the input signal $\bar u$ is continuously differentiable for $0 \leq t < T$. This is required such that (\ref{difeqa}) is well defined over this interval. However, the following corollary of Theorem \ref{thmctwill} extends this result to include piecewise differentiable inputs $\bar u$. Here, we use the notation $z(t^-) := \lim_{\varepsilon \rightarrow 0} z(t - \varepsilon)$ for any signal $z(t)$ and a scalar $\varepsilon > 0$.

\begin{cor}
	Let the conditions in Theorem \ref{thmctwill} hold. Moreover, consider a set of instants $0 \leq T_j < T$, $T_j < T_{j+1}$, $j=0,\ldots,M$, with $T_0 := 0$, $T_{M+1} := T$. Then, any signals $\bar u : [0,T] \rightarrow \mathbb{R}^m$, $\bar y : [0,T] \rightarrow \mathbb{R}^p$, where $\bar u$ is continuously differentiable for all $t \in [0,T]$ with $t \neq T_j$, and continuously differentiable from the right at the instants $t = T_j$, are an input-output trajectory of (\ref{linsys}) corresponding to some initial condition $\bar x(0)$, if and only if there exists a time-varying vector $\alpha(t)$ such that the equations (\ref{difeqa}), (\ref{inicoa}) and (\ref{ctwillem}) hold piecewise in the intervals $T_{j} \leq t < T_{j+1}$, and
	\begin{multline}
		\left[ \begin{array}{c}
			H_{1,T}(u_{[0,N-1]}(T_j)) \\ H_{1,T}(x_{[0,N-1]}(T_j))
		\end{array} \right] \alpha(T_j) = \\
		\left[ \begin{array}{c}
			\bar{u}(T_j) \\ H_{1,T}(x_{[0,N-1]}(T_j^-)) \alpha(T_j^-)
		\end{array} \right]
	\label{iniint}
	\end{multline}
for $j=1,\ldots,M$. 
\end{cor}
\begin{proof}
	Since $\bar u$ is continuosly differentiable from the right, the steps in the proof of Theorem \ref{thmctwill} hold in each interval $T_{j} \leq t < T_{j+1}$, where (\ref{iniint}) provides the initial conditions for $\alpha(t)$ in this interval. In particular, the second row of (\ref{iniint}) ensures that $\alpha(t)$ is such that $\bar{x}(t)$ is also continuous at the time instances $T_j$.
\end{proof}

\subsection{On system state availability}
In the CT version of Willems' lemma presented in Theorem~\ref{thmctwill}, the vector $\alpha(t)$ is time-varying and must solve the differential equation (\ref{difeqa})-(\ref{inicoa}). Some of the existing analysis and control results for discrete-time systems using Willems' lemma assume that state measurements are available (compare, e.g., \cite{BerberichKocSchAll2020,DorflerTesDe2021,TuranFer2022,WolffLopMul2021} and Sections IV and V in \cite{DePersisTes2020}), while others directly work with input/output data only. In particular, when using the DT Willems' lemma for the prediction of system trajectories such as in the data-based simulation problem \cite{MarkovskyRap2008,BerberichAll2020} or in data-based MPC \cite{CoulsonLygDoe2019,BerberichKoeMueAll2021}, a suitable vector $\alpha$ can be computed using input/output data only. In contrast, computing for $\alpha(t)$ by solving (\ref{difeqa})-(\ref{inicoa}) requires that an a priori collected state sequence is available, similar to \cite{DePersisTes2020,BerberichKocSchAll2020,DorflerTesDe2021,TuranFer2022,WolffLopMul2021}.  Examining alternatives to (\ref{difeqa})-(\ref{inicoa}) that do not require state measurements (but input/output measurements only) is a subject of future research.

\subsection{Practical considerations}
\label{subsecprac}
Generating continuous-time input-output system trajectories requires the solution of (\ref{difeqa})-(\ref{inicoa}) and (\ref{ctwillem}) for all $0 \leq t < T$. Naturally, to apply these procedures using digital devices, only samples of the CT signals can be used. This means that the trajectories generated in practice using the method proposed here will be an approximation of the true system trajectories, in the same sense that any computer simulation of a continuous-time system provides only an approximation. Still, this approximation takes into account the fact that the input signal is not held constant between samples, as would be the case when using the discrete-time Willems' lemma.

Another advantage of our method over the use of the DT Willems' lemma is as follows. Assume that a sample of data from the CT signals is collected every $\delta$ time units. Then, to generate trajectories of length $T$ using the DT Willems' lemma (\ref{dtwillem}), $L$ takes the value $L=T/\delta$. Moreover, the total amount of a priori collected data samples is $M = NT/\delta$. Since the matrix on the left-hand side of (\ref{dtwillem}) has dimension $L(m+p) \times (M-L+1)$, a finer sampling of the CT signals (i.e., a smaller $\delta$) implies a considerable increase in the dimension of the matrix. In contrast, the dimensions of the matrices in (\ref{difeqa})-(\ref{inicoa}) and (\ref{ctwillem}) are fixed given the value of $N$ (see (\ref{hankrow})). Using our method, a decrease in $\delta$ implies only a larger \emph{amount} of matrices computed in the interval $0 \leq t < T$ (one for each value $t=i \delta$, $i=0,\ldots,T/\delta-1$).

Finally, notice that the samples collected from the continuous-time signals allow for a numerical solution of the differential equation (\ref{difeqa})-(\ref{inicoa}). Software packages like Matlab, for example, can solve the equivalent ODE (\ref{odea}) from sampled data using built-in functions.

\section{CONCLUSIONS}
\label{secconc}

In this paper, we showed sufficient and necessary conditions to represent the input-output trajectories of a continuous-time system using measured persistently excited data. This corresponds to a continuous-time version of the well-known Willems' lemma. Different from the discrete-time case, additional conditions on the vector of parameters $\alpha(t)$ must be satisfied. Namely, $\alpha(t)$ is time-varying and must solve the differential equation (\ref{difeqa})-(\ref{inicoa}). Using this method, input-output trajectories of a desired length $T$ can be generated, where the input $\bar u$ is piecewise differentiable.

Throughout the paper, possible future lines of research were proposed, in particular with respect to a framework including input/ouput (but no state) data, and obtaining PE data from input signals that are not piecewise constant. Additionally, a natural research direction is to use the results presented in this paper for the development and analysis of data-based controllers for continuous-time systems.


\bibliographystyle{IEEEtran}
\bibliography{IEEEabrv,dbcontrol_refs}


\end{document}